\documentclass[prb,preprint,showpacs]{revtex4}
\usepackage[dvips]{graphicx}

\newcommand{\ignore}[1]{}
\newcommand{\beq}{\begin{equation}}
\newcommand{\eeq}{\end{equation}}
\newcommand{\pd}{{\partial\over\partial t}}

\begin{document}

\title{Quantum Spinodal Phenomena}
\author{Seiji Miyashita}
\email[Corresponding author: ]{miya@spin.phys.s.u-tokyo.ac.jp},
\affiliation{Department of Physics, Graduate School of Science,
The University of Tokyo, 7-3-1 Hongo, Bunkyo-Ku, Tokyo 113-8656, Japan}
\affiliation{CREST, JST, 4-1-8 Honcho Kawaguchi, Saitama 332-0012, Japan}
\author{Hans De Raedt}
\affiliation{Department of Applied Physics,
Zernike Institute of Advanced Materials,
University of Groningen, Nijenborgh 4, NL-9747 AG Groningen, The Netherlands}
\author{Bernard Barbara}
\affiliation{Institut N\'eel, CNRS,
25 Ave. des martyrs, BP 166, 38 042 Grenoble Cedex 09, France}
\date{\today}

\begin{abstract}
We study the dynamical magnetization process in the ordered ground-state phase 
of the transverse Ising model under sweeps of magnetic field with constant velocities.
In the case of very slow sweeps and for small systems studied previously 
(Phys. Rev. B 56, 11761 (1997)), 
non-adiabatic transitions at avoided level-crossing points give the dominant contribution 
to the shape of magnetization process. 
In contrast, in the ordered phase of this model and for fast sweeps, 
we find significant, size-independent jumps in the magnetization process. 
We study this phenomenon in analogy to the spinodal decomposition in classical ordered state
and investigate its properties and its dependence on the system parameters. 
An attempt to understand the magnetization dynamics under field sweep 
in terms of the energy-level structure is made.
We discuss a microscopic mechanism of magnetization dynamics from a viewpoint of 
local cluster flips and
show that this provides a picture that explains the size independence.
The magnetization dynamics in the fast-sweep regime is studied by perturbation theory
and we introduce a perturbation scheme based on interacting Landau-Zener type processes
to describe the local cluster flip dynamics. 
\end{abstract}


\maketitle

\section {Introduction}

It is well-known that non-adiabatic transitions among adiabatic eigenstates
take place when an external field is swept with finite velocity~\cite{LZ,LZmiya}.
In particular, at avoided level-crossing points strong non-adiabatic transitions 
occur, causing a step-wise magnetization process~\cite{LZ1997}.

In so-called single molecular magnets~\cite{SMM}, 
the energy level diagram consists of discrete levels because the molecules 
contain only small number of magnetic ions
and hence the quantum dynamics plays important roles.
In particular, in the easy-axis large spin molecules 
such as Mn$_{12}$ and Fe$_8$, step-wise magnetization processes 
have been found and they are attributed to the 
adiabatic change, that is the quantum tunneling at the avoided level-crossing points, 
and are called resonant tunneling phenomena~\cite{SMM1}.
The Landau-Zener mechanism also causes various interesting 
magnetization loops in field cycling processes~\cite{SMM2}.

The amount of the change 
of the magnetization at a step in the magnetization process
is governed by the Landau-Zener mechanism and
depends significantly on the energy gap at the crossing. 
This dependence has played an important role in the study of single-molecule magnets.
Observations of the gap have been  done on isolated magnetic molecules~\cite{LZSexperiments}.

The quantum dynamics of systems of strongly interacting systems which show quantum phase transitions 
is also of much contemporary interest.
As far as static properties are concerned, 
the action in the path-integral representation of a $d$-dimensional quantum system 
maps onto the partition function of 
a $(d+1)$-dimensional classical model, 
which is the key ingredient of the quantum Monte Carlo simulation.\cite{QMC}
From this mapping, it follows that
the critical properties of the ground state of the $d$-dimensional 
quantum system are the same as those of equilibrium state of 
the $(d+1)$-dimensional classical model, quantum fluctuations playing
the role of the thermal fluctuations at finite temperatures. 

However, from a view point of dynamics, the nature of the quantum and thermal
fluctuations are not necessarily the same. 
Thus, it is of interest to study dynamical aspects of quantum critical phenomena. 
As a typical model showing quantum critical phenomena,
in the present work we adopt the one-dimensional transverse Ising model~\cite{TI}.

Recently, interesting properties of molecular chains which are modeled by the transverse Ising model 
with large spins have been reported~\cite{Yamashita}.
However, in this paper, we concentrate ourselves in systems of $S=1/2$. 
The dynamics of the transverse Ising model plays important roles in the study of the quantum annealing
in which the quantum fluctuations due to the transverse field are used to survey the ground state in a complex system~\cite{QA}. 
The dynamics of domain growth under the sweep of the transverse field 
through the critical point has been studied related to the Kibble-Zurek mechanism~\cite{Hxsweep,KZ}.

In this paper, we study the hysteresis behavior as a function 
of the external field in the ordered state by performing simulations 
of pure quantum dynamics, that is by solving the time-dependent Schr\"odinger equation.\cite{HDRQD}
This gives us numerically exact results of 
the dynamical magnetization process of the transverse Ising model
under sweeps of magnetic field with constant velocities.

Previously we have studied time evolution of magnetization of the transverse Ising model 
from a view point of Landau-Zener transition, sweeping the field slowly 
and finding transitions at each avoided level crossing point.\cite{LZ1997}
However, for fast sweeps the transition at zero field $H_z=0$ disappears and
the magnetization does not change even after the field reverses. 
The magnetization remains in the direction opposite to the external field for a while, 
and when the magnetic field reaches an certain value, the magnetization
suddenly changes to the direction of the field.
This sudden change is also found for very slow sweeps at the level crossing point. 
However, the present case has the following two differences: 
(1) the switching field does not necessarily corresponds to a level crossing, and 
(2) in all cases the changes are independent of the size $L$ of the system.
This sudden change resembles the change of magnetization at the coercive field in
the hysteresis loop of ferromagnetic systems, where it is called spinodal decomposition.
Therefore, we will call the phenomenon that we observe in the quantum system a "quantum spinodal decomposition"
and the field ``quantum spinodal point'' $H_{\rm SP}$.
We study the dependence of $H_{\rm SP}$ on the transverse field $H_x$,
and also study the sweep-velocity dependence of $H_{\rm SP}$.

As in the cases of the single molecular magnets, it should be possible to understand
the dynamics of the magnetization in terms of the energy levels as a function of field.
However, because the structure of the energy-level diagram strongly depends on the size of the system,
it is difficult to explain the size-independent property of the quantum spinodal 
decomposition from the energy-level structure only.
In the case of much faster sweeps, we find almost perfect 
size-independent magnetization processes. 
We also find a peculiar dependence of magnetization on the field in the case of weak transverse fields.
These processes can be understood from the energy-level diagram for local flips of spins, 
but not from the energy diagram of the total system.

In this paper, we attempt to understand the microscopic mechanism that gives rise to this size
independent dynamics. We introduce a perturbation scheme for fast sweeps,
regarding the fast sweeping field term as the unperturbed system and treating
the interaction term as the perturbation.
From this viewpoint, we investigate fundamental, spatially local time-evolutions which 
yield the size-independent response to the sweep procedure.
In particular, we propose a perturbation scheme in terms of independent Landau-Zener systems,
each of which consists of a spin in a transverse and sweeping field.
A system consisting of locally interacting Landau-Zener systems explains well the
magnetization dynamics for fast sweeps.


\section{Model}
We study characteristics of dynamics of the order parameter 
of the one-dimensional transverse-Ising model with periodic 
boundary condition under a sweeping field.\cite{TI}
The Hamiltonian of the system is given by
\beq
{\cal H}(t)=-J\sum_i \sigma_i^z\sigma_{i+1}^z-H_x\sum_i\sigma_i^x
         -H_z(t)\sum_i\sigma_i^z,
\label{ham}
\eeq
where $\sigma_i^x$ and $\sigma_i^z$ are the $x$ and $z$ components of the
Pauli matrix, respectively. Hereafter, we take $J$ as a unit of the energy.
The order parameter is the $z$ component of the magnetization
\beq
M^z=\sum_i\sigma_i^z.
\eeq
We study dynamics of the order parameter of the model, i.e.,
the time dependence of the magnetization under the time dependent
field $H_z(t)$ 
\beq
\langle M^z\rangle= \langle\Psi(t)|M^z|\Psi(t)\rangle,
\eeq
where $|\Psi(t)\rangle$ is a time dependent wavefunction given by
the Schr\"odinger equation
\beq
i\hbar{\partial\over\partial t}|\Psi(t)\rangle={\cal H}(t) |\Psi(t)\rangle.
\eeq
In the present paper we study the case of linear sweep of the field
\beq
H_z(t)=-H_0+ct,
\eeq
where $-H_0$ is an initial magnetic field. In the present paper, we set $H_0/J=1$, and
$c$ is the speed of the sweep. We use a unit where $\hbar=1$.

In the case $H_z=0$, the model shows an order-disorder phase transition
as a function of $H_x$.
The transition point is given by $H_x^{\rm c} = J$.
In the ordered phase ($H_x < J$), the system has a spontaneous magnetization
$m_{\rm s}$:
\beq m_{\rm s}=\lim_{H_z\rightarrow +0}\lim_{L\rightarrow \infty}
\langle G(0)|M^z|G(0)\rangle,
\eeq
where $|G(0)\rangle$ is the ground state of the model with $H_z=0$.
Therefore, the ground state is twofold degenerate with symmetry-broken magnetization, 
while the ground state is unique when $H_x > J$.
Because of these twofold symmetry-broken ground states, the magnetization
changes discontinuously at $H_z=0$.

In a finite system $L < \infty$, this degeneracy is resolved by
the quantum mixing (tunneling effect) and a small gap opens
at $H_z=0$. This gap becomes small exponentially with $L$ as shown 
in Appendix A. Therefore, the change of the magnetization 
becomes sharper as $L$ increases.
Dynamical realization of this change by field sweeping 
becomes increasingly difficult with $L$. This phenomenon corresponds to the
existence of metastable state.

The energy-level diagram becomes complicated when $L$ increase.
However, as shown below, when $c$ is large 
the system shows a size-independent magnetization dynamics
which is not easily understood in terms of the energy-level diagram.
In this paper, we focus on the regime of moderate to large sweep velocities.


\section{Energy structure}

\begin{figure}[t]
\begin{center}
\mbox{
\includegraphics[width=8cm]{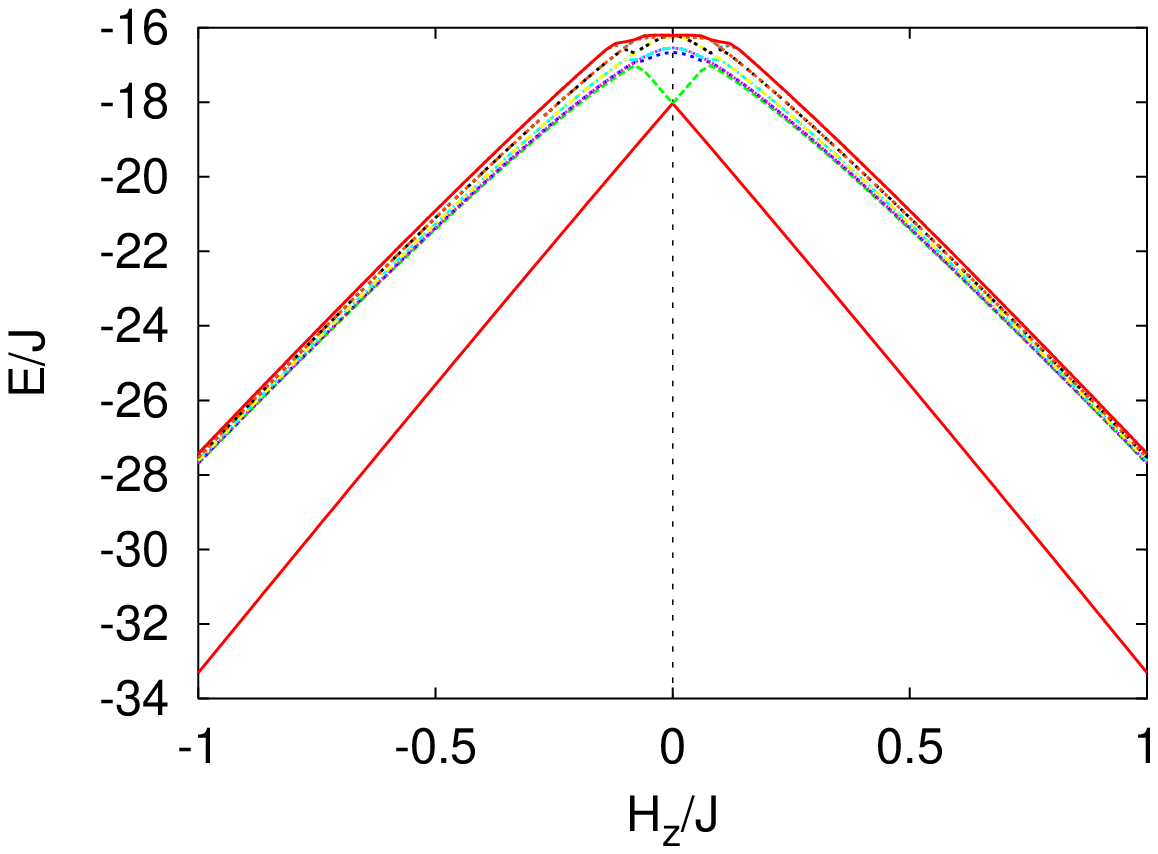}
\includegraphics[width=8cm]{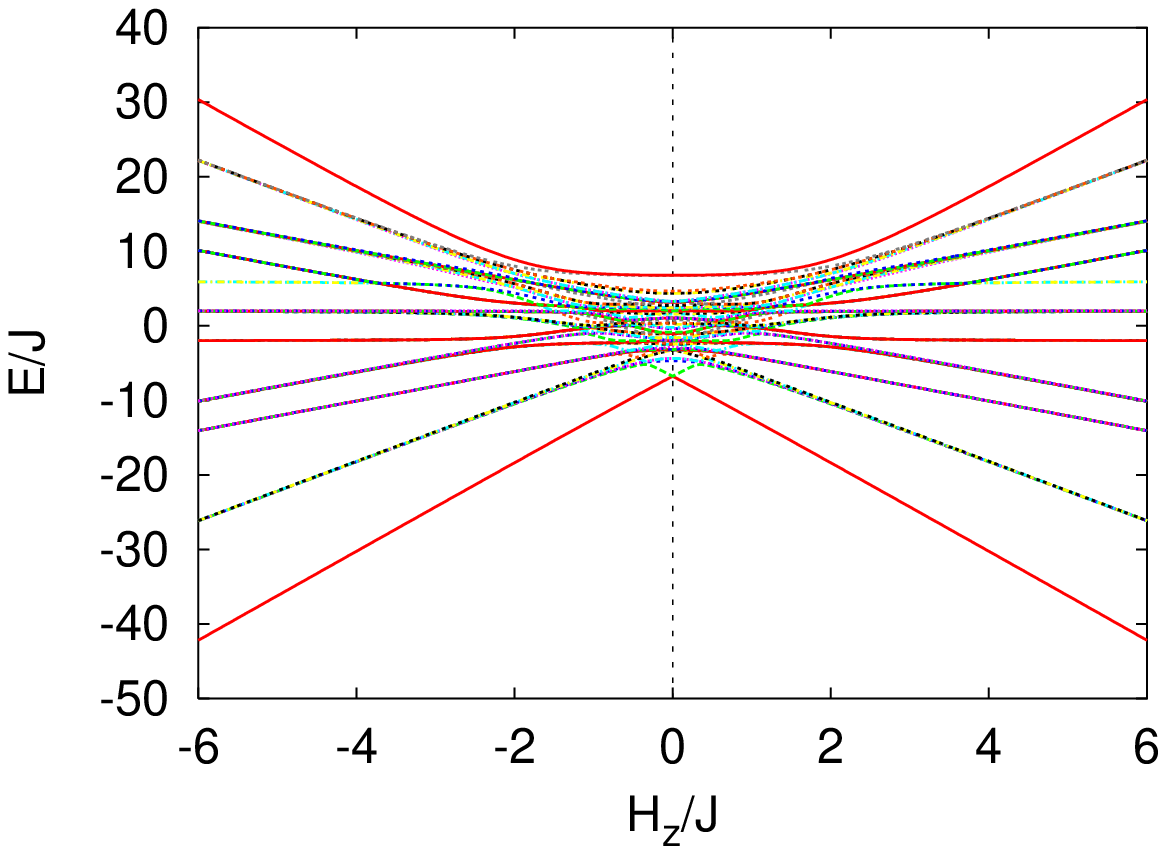}
}
\caption{(Color online)
Typical energy-level diagrams of model Eq.~(1).
Left: full spectrum for $L=6,H_x=0.7$.
Right: A few low-energy states for $L=16,H_x=0.7$.
}
\label{engHx07L12}
\end{center}
\end{figure}

In Fig.~\ref{engHx07L12}(left), we present 
an energy-level diagram for $L=6$ and $H_x=0.7$.
We plot all energy levels as a function of $H_z$.
We find that the energy levels show a linear dependence at large fields,
where quantum fluctuations due to $H_x$ have little effect.
The levels are mixed in the region $-3<H_z/J <3$ where 
the energy levels come close and are mixed by the transverse field.
The isolated two lowest energy levels are located under 
a densely mixed area, which represent the ordered states with $M=L$ and $-L$,
and they cross at $H_z=0$ with a small gap $\Delta E_1$, 
reflecting the tunneling between the symmetry broken states.
The gap $\Delta E_1$ is so small that one cannot see it in Fig.~\ref{engHx07L12}(left).
After the crossing, these states join the densely mixed area.
In Fig.~\ref{engHx07L12}(right), we show the energy levels for $L=16$ where 
we plot only energies of a few low-energy states. In this figure,
we also find the above mentioned characteristic structure of two lowest energy levels. 

Let us point out a few more characteristic features of the energy-level diagram.
A finite gap $\Delta E_2$ exists between the crossing point of the low-lying lines 
and the densely populated region of excited states. 
The $d$-dimensional Ising model in a transverse field is closely related to
the transfer matrix of $(d+1)$-dimensional Ising model.
From this analogy, we associate $\Delta E_1$ to symmetry breaking phenomena.
When symmetry breaking takes place, the two largest eigenvalues of the transfer 
matrix of the model become almost degenerate. 
The energy gap corresponds to the tunneling through the free-energy barrier between the two ordered states
and vanishes exponentially with the system size.
On the other hand, $\Delta E_2$ is related to the correlation length of the fluctuation of
anti-parallel domains in the order state.
The correlation length is finite at a given temperature in the ordered state and is almost size-independent.
At $H_z=0$ we can calculate eigenenergies analytically, 
and we can explicitly confirm that
$\Delta E_1$ vanishes exponentially with $L$
and that $\Delta E_2$ is almost constant as a function of the size. 
The dependencies of the energy gaps at $H_z=0$ are discussed in Appendix A.

For large $H_z$, the slopes of the low lying isolated lines are $\pm L$ 
because they represent the states with $M=\pm L$. 
Thus, the field at which the lines merge in the area of densely populated 
excited states is given by
\beq
H_z\simeq {\Delta E_2\over L}\equiv H_z^{\ast}(L).
\eeq
At this point, the magnetization shows a jump when the speed of the sweep is very slow.\cite{LZ1997}

However, as we will see in the following sections, the dynamical magnetization does not show 
any significant change at this field value when the sweeping field is fast. 
Another type of jump will occur that we called quantum spinodal jump or quantum spinodal transition.

\section{Evolution of the magnetization for fast sweeps of the field}


\subsection{Quantum spinodal decomposition}

\begin{figure}
\begin{center}
\mbox{
\includegraphics[clip,width=8cm]{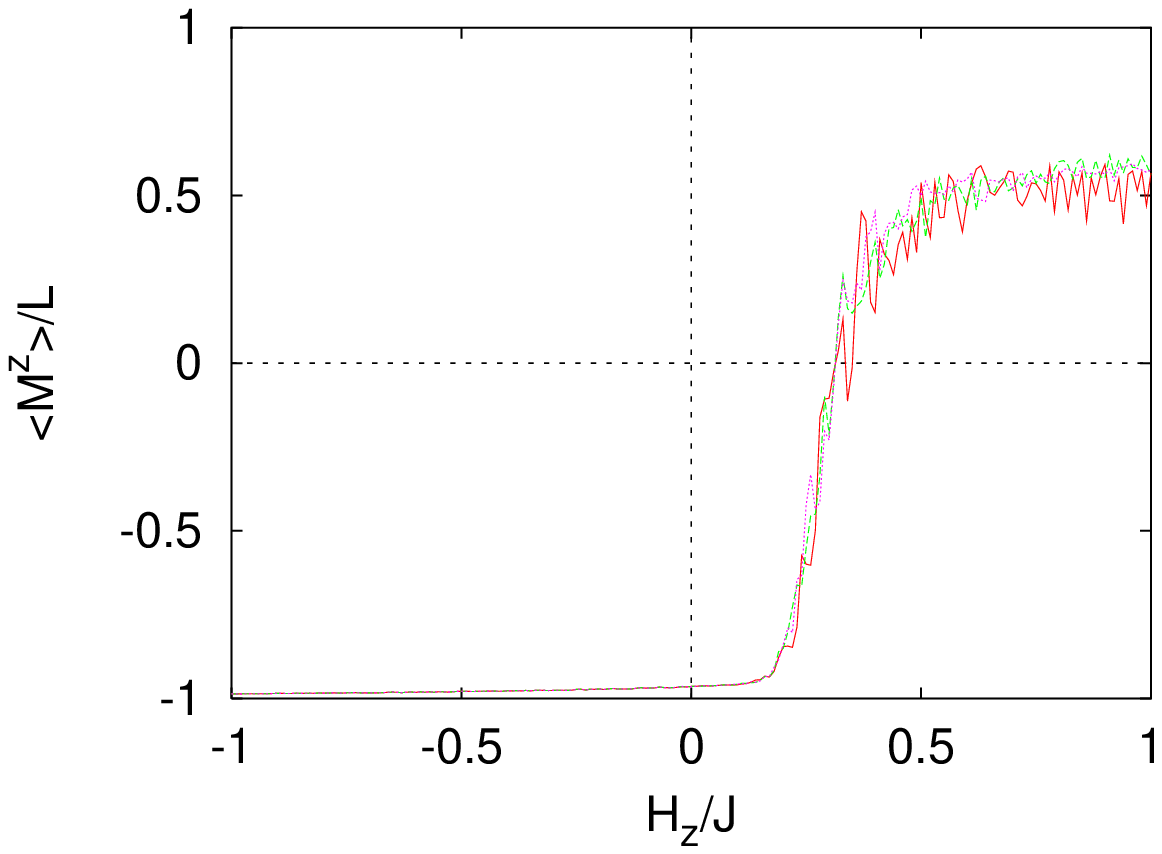}
\includegraphics[clip,width=8cm]{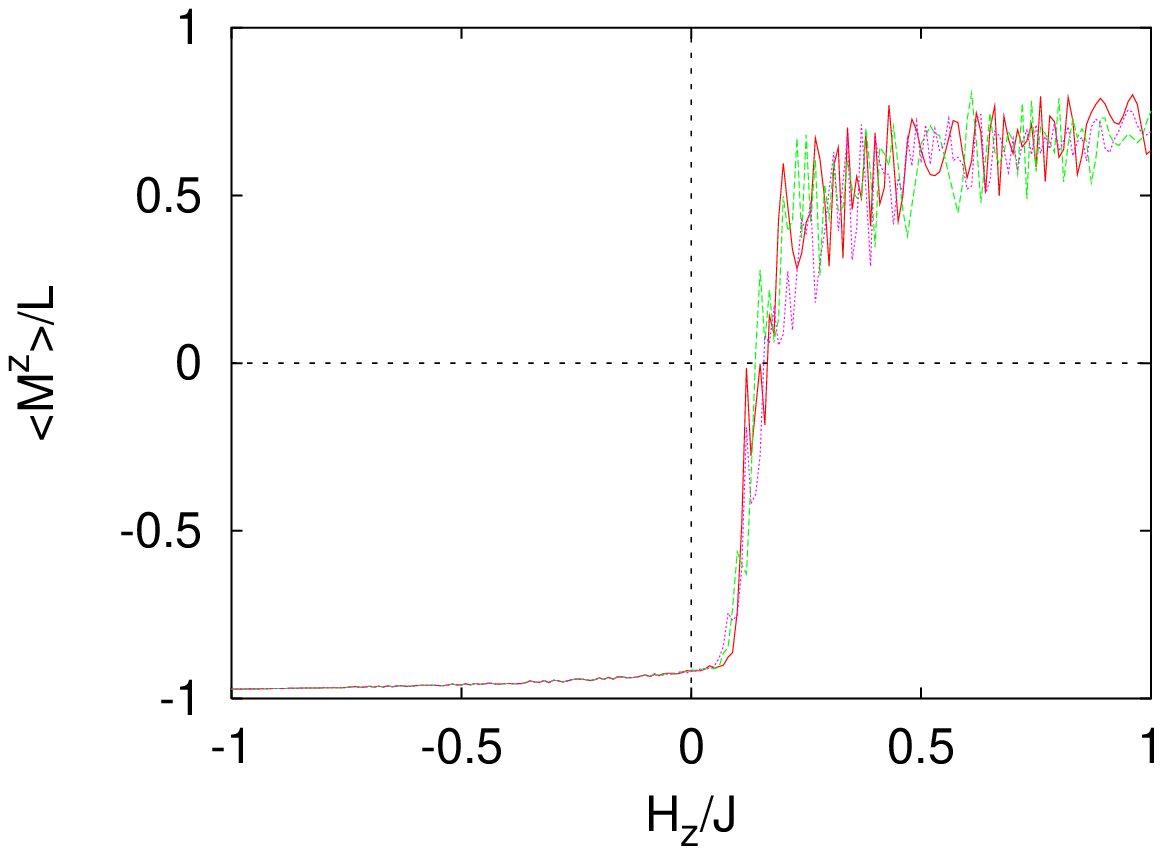}
}
\caption{(color online)
Left: Magnetization $M^z(t)$ as a function of $H_z(t)$ for $c=0.001$,
$H_x=0.5$ and various system sizes.
Solid (red) line: $L=12$;
Dashed (green) line: $L=14$;
Dotted (magenta) line: $L=16$;
Right: Same as left except that $H_x=0.7$.
}
\label{sweepHx05L12c001}
\end{center}
\end{figure}

When we sweep the magnetic field from $H_z=-1$ to $H_z=1$, 
the magnetization shows a rapid increase to a positive value. 
In Fig.~\ref{sweepHx05L12c001}, we depict examples of dynamics of the
magnetization as a function of time for a sweeping velocity $c=0.001$. 
Because $H_z(t)=-H_0+ct$, $H_z$ also represents time. 

The magnetization stays at a negative value until a certain field strength is reached.
The system can be regarded as being in a metastable state. Then, the magnetization 
changes significantly towards the direction of the field in a single continuous jump,
the magnetization processes $M_z(t)$ depending very weakly on the system size. 
In the classical ordered state, we know a similar behavior. Namely,
at the coercive field (at the edge of the hysteresis), the magnetization
relaxes very fast and the relaxation time does not depend on the size.
Thus, we may make an analogy to the spinodal decomposition phenomena.
We call the phenomenon that we observe in the quantum system ``quantum spinodal decomposition'' and
we call the field at which the magnetization changes $H_{SP}$.
It should be noted that the spinodal decomposition corresponds to the fact that
the size of the critical nuclei 
becomes of the order one. If the size of the
critical nuclei is larger than the size of the particle as 
in the case of nanoparticles, 
the critical field of the sudden magnetization reversal, 
which is also a kind of spinodal decomposition,
strongly depends on the size.

First, let us attempt to understand this dynamics from 
the view point of the energy diagram. 
As we mentioned in the previous section,
the low-lying levels of $M=\pm L$ merge with the continuum at $H_z=H_z^{\ast}$. 
Thus, we expect that at this point the magnetization begins to change
because the states with $M=\pm L$ begin to cross other states.
In fact, in earlier work, we found stepwise magnetization processes 
at avoided level-crossings in very slow sweeps, each of which
could be analyzed in terms of successive Landau-Zener crossings.\cite{LZ1997}

From Fig.~\ref{sweepHx05L12c001}(left), we find
that the sharp change of $M_z(t)$ starts at $H_z= 0.2\sim0.25$,
which is much larger than $H_z^\ast(L)$. 
We estimate $H_z^\ast(12)\simeq 0.18$, and for larger lattices $H_z^\ast(L)$ is even smaller for larger lattices.
Moreover, it should be noted that the magnetization processes display almost no size-dependence. 
In Fig.~\ref{sweepHx05L12c001}(right), which shows $M_z(t)$ for $H_x=0.7$,
we also find that the magnetization processes $M_z(t)$ for all sizes $L$ 
are very similar. Here $H_{SP} \approx 0.11$ is again significantly larger than $H_z^\ast(L)$  
(for $L=16$ and $\Delta E_2\approx1.4$ in the case $H_x=0.7$, and hence $H_z^\ast(14)\simeq 0.09$).
This observation is in conflict with the picture 
based on the structure of the energy level diagram given earlier.

\begin{figure}
\begin{center}
\includegraphics[width=14cm]{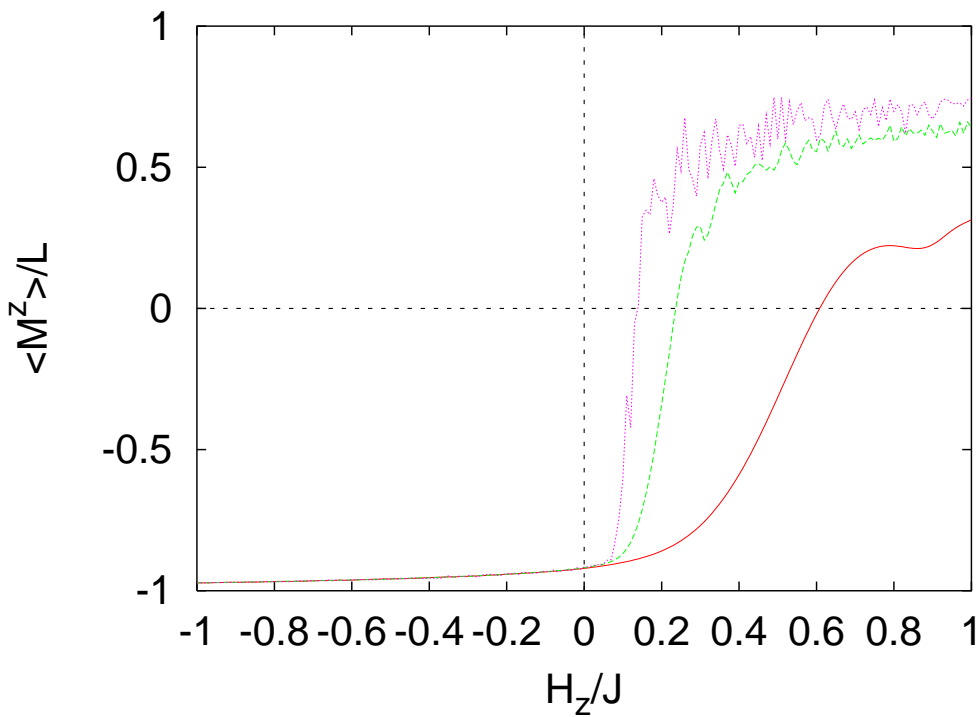}
\caption{
Magnetization $M^z(t)$ as a function of $H_z(t)$ for $L=20$,
$H_x=0.7$ and various sweep velocities.
Solid (red) line: $c=0.1$;
Dashed (green) line: $c=0.01$;
Dotted (magenta) line: $c=0.001$.
}
\label{fig3}
\end{center}
\end{figure}

In Fig.~\ref{fig3}, we present an example of sweep-velocity dependence for a system with $L=20$ 
(results of other sizes are not shown).
The magnetization processes show strong dependence on the sweep velocity 
$c$, as expected. However, for fixed $c$, there is little dependence on $L$
(results of other sizes are not shown).   

\begin{figure}
\begin{center}
\includegraphics[clip,width=14cm]{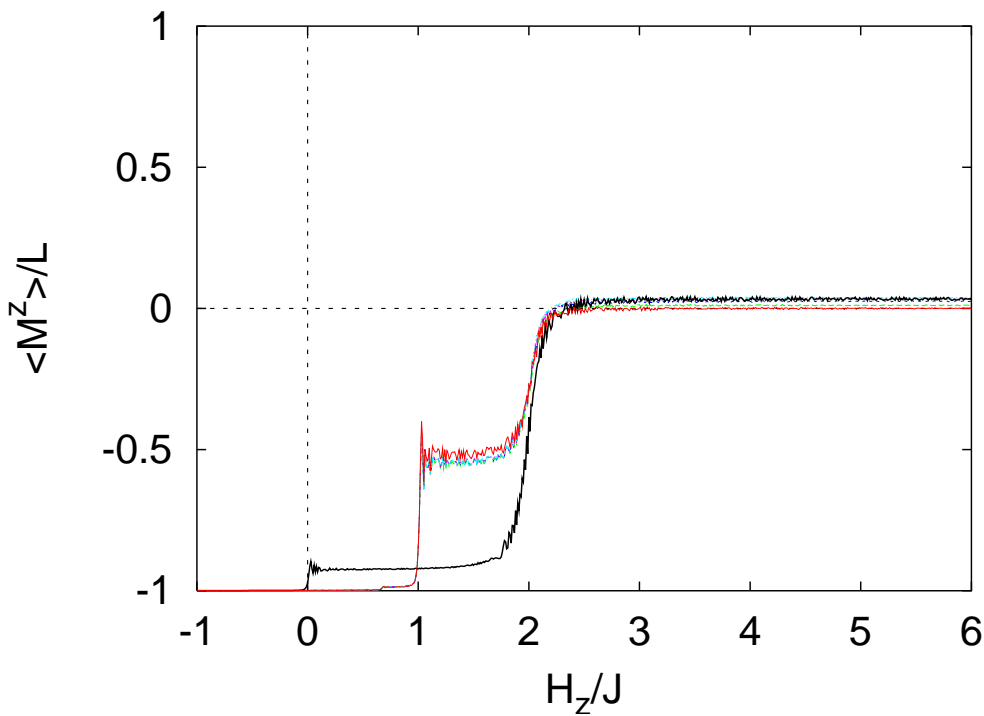}
\caption{Magnetization $M^z(t)$ as a function of $H_z(t)$ for
$H_x=0.1$, $c=0.001$, and various system sizes.
Solid (black) line: $L=2$;
Solid (red) line: $L=4$;
Long dashed (green) line: $L=6$;
Dashed (magenta) line: $L=8$;
Dotted (red) line: $L=10$;
Dashed dotted (blue) line: $L=12$;
}
\label{LZ-perturbation01}
\end{center}
\end{figure}

We have found the characteristic change in the cases of 
relatively large quantum fluctuations, i.e.
$H_x=0.5$ and 0.7.
The size-independence indicates that the change occurs locally. 
When $H_x$ is small, the quantum fluctuations are weak and local flips of clusters
consisting of small number of spins become dominant.
In Fig.~\ref{LZ-perturbation01}, $M^z(t)$ for $H_x=0.1$ is shown, where  
a peculiar sequence of jumps is found. 
It is almost independent of the system size (except for $L=2$). 
Before the large jump of the magnetization at $H_z/J=1$, there is a small but
non-zero precursor jump around $H_z\simeq 2/3$. 
After these jumps, the magnetization shows a plateau of $M^z(t)/L\approx-1/2$
until the smooth crossover to the saturated value takes place around $H_z/J=2$.
The value $H_z/J=2$ corresponds to the spinodal point of 
the corresponding classical model.

The positions of these jumps can be understood from 
the viewpoint of local "cluster" flips.
Let us consider a single spin flip, that is, 
a flip from the state with all spins $|------- \cdots\rangle$ 
to a state $|---+--- \cdots\rangle$.
The diabatic energies of these states are $E_0=-LJ+LH_z$ and $E_1=-(L-4)J-(L-2)H_z$,
respectively. Thus, the crossing of these states occurs at $H_z^{(1)}=4J/2=2J$.
The transition probability due to the transverse field $H_x$ at this crossing is
proportional to $H_x^2$, because the matrix element for a single flip is
proportional to $H_x$.

If we consider a collective flip of a connected cluster of $m$ spins,
the diabatic energy of this state is 
\beq
E_m=-(L-4)J+(L-2m)H_z,
\label{flipm}
\eeq 
and thus, the crossing of the states occurs at 
\beq
H_z^{(m)}=4J/2m=2J/m.
\label{xxx}
\eeq
For $m=2, 3, \ldots$ we have $H_z^{(m)}=1, 2/3, \ldots$ respectively.
These values do not depend on $L$.
%
It should be noted that for the system $L=2$, the 2-spin cluster (m=2)
surrounded by $+$ spins can not be realized, and no jump appears at $H_z/J=1$. 

The matrix element for the $m$-spin cluster flip is proportional to $H_x^m$ 
(see Appendix A). Therefore, for small $H_x$, 
only the flips with small values of $m$ are appreciable.
In the case of $H_x/J=0.1$ for $c=0.001$, jumps for $m\le 2$ are observed.
The change of the magnetization of each spin is given by a perturbation series 
and is independent of $L$ as shown in Appendix B.
These local flips may correspond to the nucleation in classical dynamics in
metastable state.  

If $c$ becomes small or $H_x$ becomes large, 
contributions from large values of $m$ become relevant.
Then, magnetization process consists of many jumps, and amount of the change
becomes large. 
But, as long as the perturbation series converges, 
we have a size-independent magnetization process,
as shown in Fig.~\ref{sweepHx05L12c001}.
This sharp and size-independent nature is consistent with the property 
of the classical spinodal decomposition. 

In the classical system,
the magnetization relaxes to its equilibrium value
at the spinodal decomposition point.
In contrast, for pure quantum dynamics, 
the magnetization of the state does not change for adiabatic motion
along a particular energy level.
Only if we include an effect of contact with the thermal bath, relaxation
to the ground state takes place~\cite{keiji1999}.

\subsubsection{Phase diagram}

\begin{figure}
\includegraphics[clip,width=10cm]{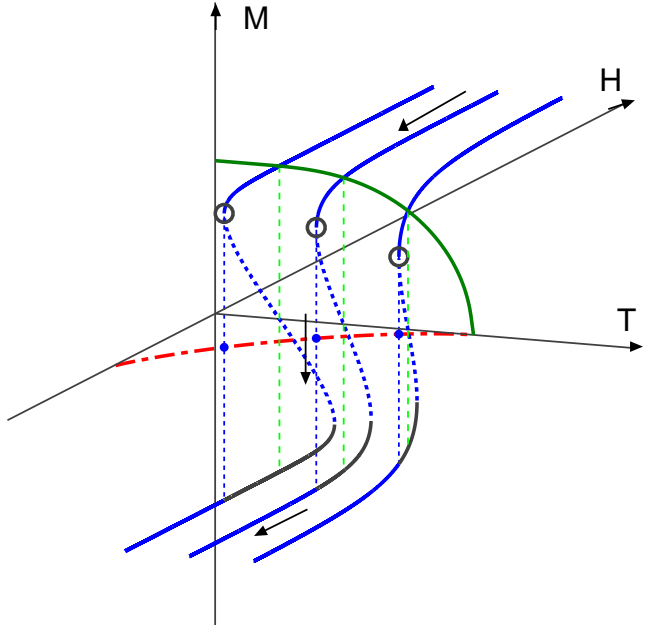} 
\caption{(color online)
Schematic picture of the magnetization $M$ as a function 
of field $H$ below the critical temperature.
Open circles denotes the spinodal decomposition points.
The (red) dash-dotted curve in the $H$--$T$ plane shows $H_{\rm SP}(T)$ as given by Eq.~(\ref{MFHSP}).
}
\label{MHxHz}
\end{figure}

In Fig.~\ref{MHxHz}, we give a schematic picture of the order parameter $M$ 
as a function of the temperature $T$ and the field $H_z$ 
in the thermal phase transition of a ferromagnetic system. 
The overhanging structure signals the existence of the metastable state. 
The spinodal point is at the edge of the metastable branch.
In this figure, the magnetic field is swept from positive to negative, and
the metastable positive magnetization jump down to the equilibrium
value at $H_{\rm SP}(T)$.

In a mean field theory for the magnetic phase transition at a finite temperature,
the spinodal point is given by
\beq
{H}_{\rm SP}=-Jz\sqrt{1-{k_{\rm B}T\over J}}+{k_{\rm B}T\over 2}
\ln\left(
{
1+\sqrt{1-{k_{\rm B}T\over J}}\over
1-\sqrt{1-{k_{\rm B}T\over J}}}\right),
\label{MFHSP}
\eeq
where $z$ is the number of nearest neighbor sites.
We show the dependence of ${H}_{SP}$ as a function of $T$ 
by a dash-dot curve in Fig.~\ref{MHxHz}.

\begin{figure}
\begin{center}
\includegraphics[clip,width=16cm]{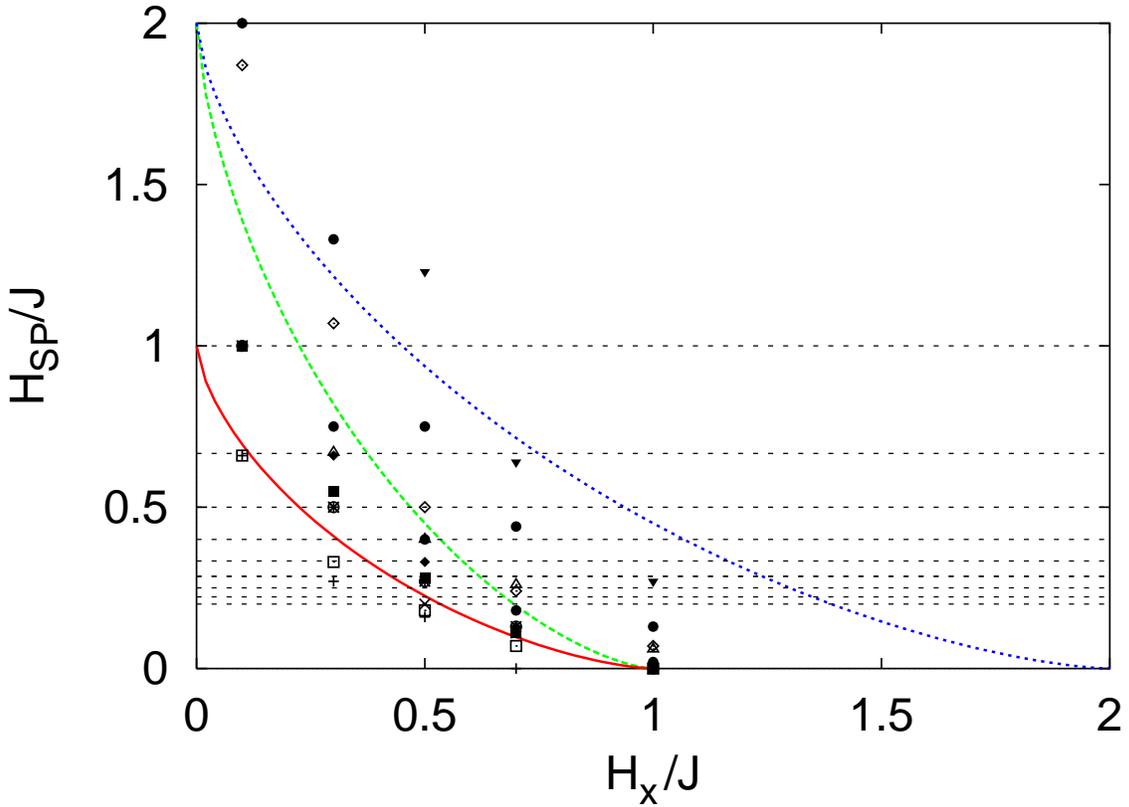}
\caption{(color online) Spinodal points $H_z^{\rm SP}$ as a function of the
quantum fluctuation $H_x$ for various sweep velocities $c$. 
The horizontal dotted lines correspond to $2/m$
for $(m=2,\ldots,10)$ (see Eq.~(\ref{xxx})). 
Plusses (1), crosses (2), and stars (3): $c=0.0001$;
Open squares (1), solid squares (2), and solid diamonds (3): $c=0.001$;
Open circles (1), bullets (2), and open diamonds (3): $c=0.01$;
Open triangles (1), solid triangles (2), and inverted solid triangles (3): $c=0.1$.
The numbers (1), (2), and (3), correspond to the field at which $M(t)$ shows a small but clear jump,
$M(t)/L=-1/2$, and $M(t)$ saturates as a function of $H_z$, respectively.
}
\label{HSPHX}
\end{center}
\end{figure}

A similar argument can be made for the classical ground state energy.
Let the $z$-component of spin be denoted by $\sigma$.
Then, the energy is expressed by
\beq
E=-J\sigma^2-H_x\sqrt{1-\sigma^2}-H\sigma.
\label{QE}
\eeq
We assume that the energy satisfies the condition
\beq
{dE\over d\sigma}=0,
\eeq
which gives
\beq
-2J\sigma+{H_x\sigma\over \sqrt{1-\sigma^2}}-H=0.
\eeq
Here, we consider the metastable state and thus we
set $H=-|H|$ for $\sigma>0$.
At the end point of metastability,
\beq
{d\sigma\over dH}=\infty \quad {\rm or} \quad {dH\over d\sigma}=0.
\eeq
This leads to
\beq
\sigma=\left(1-\left(H_x\over 2J\right)^{2/3}\right)^{1/2}.
\eeq
The end point of the metastable state is given by
\beq
H_{\rm SP}=2J\left(1-\left(H_x\over 2J\right)^{2/3}\right)^{3/2}.
\label{QSW}
\eeq
which gives $H_{\rm SP}$ as a function of $H_x$
and is shown in Fig.~\ref{HSPHX} as the long-dashed curve.
 
It is interesting to note that expression Eq.~(\ref{QSW}) is very similar 
to the well know expression of the Stoner-Wohfarth model\cite{SW}
for the reversal of a classical magnetic moment under the application of 
a magnetic field tilted with respect of the easy anisotropy axis. 
This is not surprising because, with both a longitudinal and transverse field component, 
this model can be considered as a realization of the classical spinodal transition. 
One might derive the Stoner-Wohfarth model from equation Eq.~(\ref{QE}) 
by replacing
the exchange energy parameter $J$ by the anisotropy energy constant $D$.

It should be noted that 
the critical $H_x$ is a factor of two larger than that
of the correct value $H_c^x=J$ for the one-dimensional quantum model.
This difference is due to the presence of quantum fluctuations.
Therefore, in Fig.~\ref{HSPHX} we plot Eq.(\ref{QSW})
with and without renormalized values of the fields.
The long-dashed curve denotes the case of $H_x$ scaled by 1/2, and the dashed curve
denotes the case where both $H_x$ and $H_z$ are scaled by 1/2.

As we saw in Fig.~\ref{sweepHx05L12c001}, we find a large change of magnetization
at a values of $H_z$ for each value of $H_x$, which we called $H_{\rm SP}$.
In Fig.~\ref{HSPHX}, we plot values of $H_z$ at which  
(1) $M(t)$ shows a small but clear jump,
 (2) $M(t)/L$ is equal to $-1/2$, and
 (3) $M(t)$ saturates as a function of $H_z$,
for various values of $c$.
The data show a dependence on $H_x$ that shows a similar dependence to the dotted line.
If we use other value of $c$, the values of $H_z$ change.
Although the values of $H_z$ for (1), (2) and (3) for larger values of $c$ are larger 
than those for $c=0.001$,
the values of $H_z$ for $c=0.0001$ are close to those for $c=0.001$.
They seem to saturate around the value of the dotted line,
and we may identify a sudden appearance of size independent change 
as an indication for a quantum spinodal point.
If we sweep much faster, the jumps of the magnetization becomes less clear,
as we now study in more detail.


\subsection {Very fast sweeps}

\begin{figure}[t]
\begin{center}
\includegraphics[width=14cm]{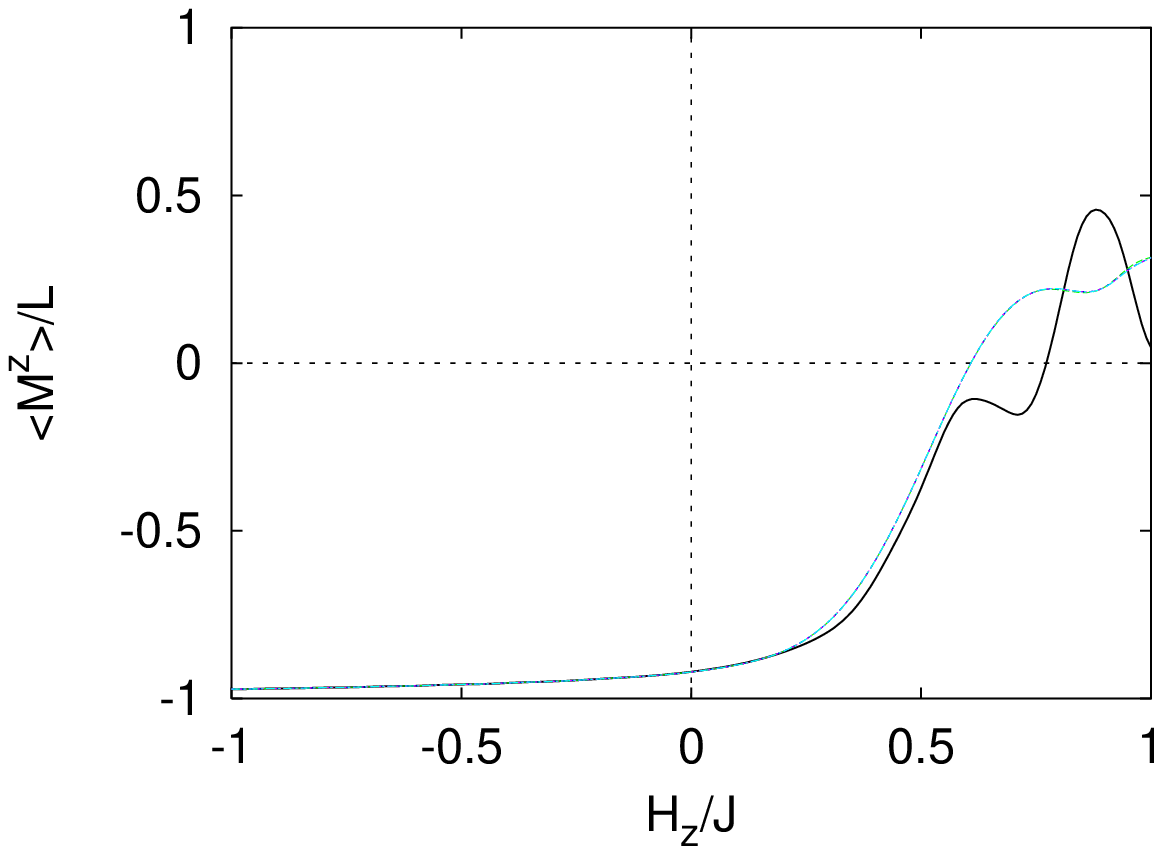}
\caption{(color online) The magnetization $M(t)$ as a function
of the sweeping field $H_z$ for 
$H_x=0.7$, $c=0.1$, and various system sizes.
Solid (black) line: $L=6$;
Solid (red) line: $L=14$;
Long dashed (green) line: $L=16$;
Dashed (magenta) line: $L=18$;
Dotted (red) line: $L=20$;
Dashed dotted (blue) line: $L=12$.
Except for $L=6$, all other curves overlap, indicating that
for sufficiently large systems, 
the dependence on $L$ is very weak.}
\label{overlap}
\end{center}
\end{figure}
For a fast sweep $c=0.1$, the magnetization processes for different sizes
almost overlap each other, see Fig.~\ref{overlap}.
The data for $L=14$, 16, and 20 are hard to distinguish.
This almost perfect overlap is rather surprising from the viewpoint 
of the structure of energy-level diagram.
The data for $L=6$ deviates from the others. This fact indicates that
for these parameters $(H_x=0.7, c=0.1)$ the relevant size of the cluster
($m$ in Eq.~(\ref{flipm})) is larger than 6 but smaller than 14.

\begin{figure}[t]
\begin{center}
\mbox{
\includegraphics[width=8cm]{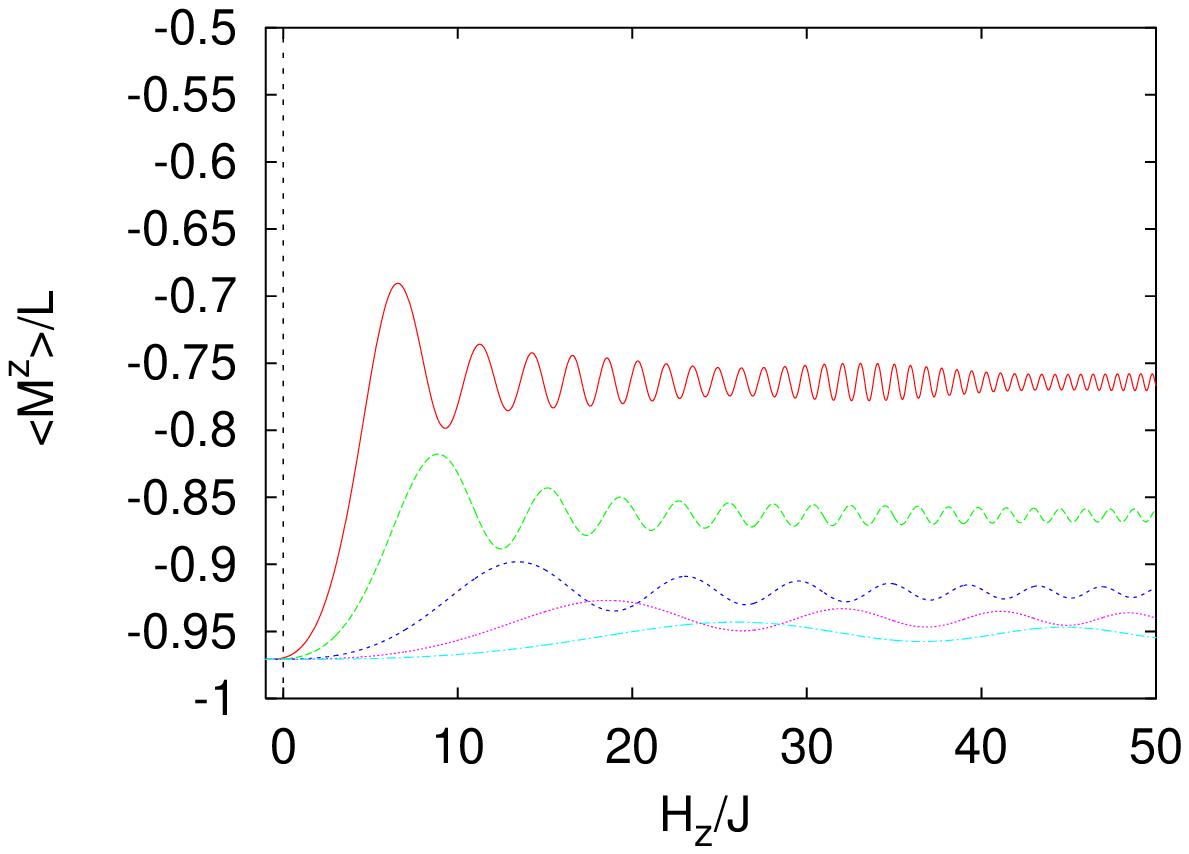}
\includegraphics[width=8cm]{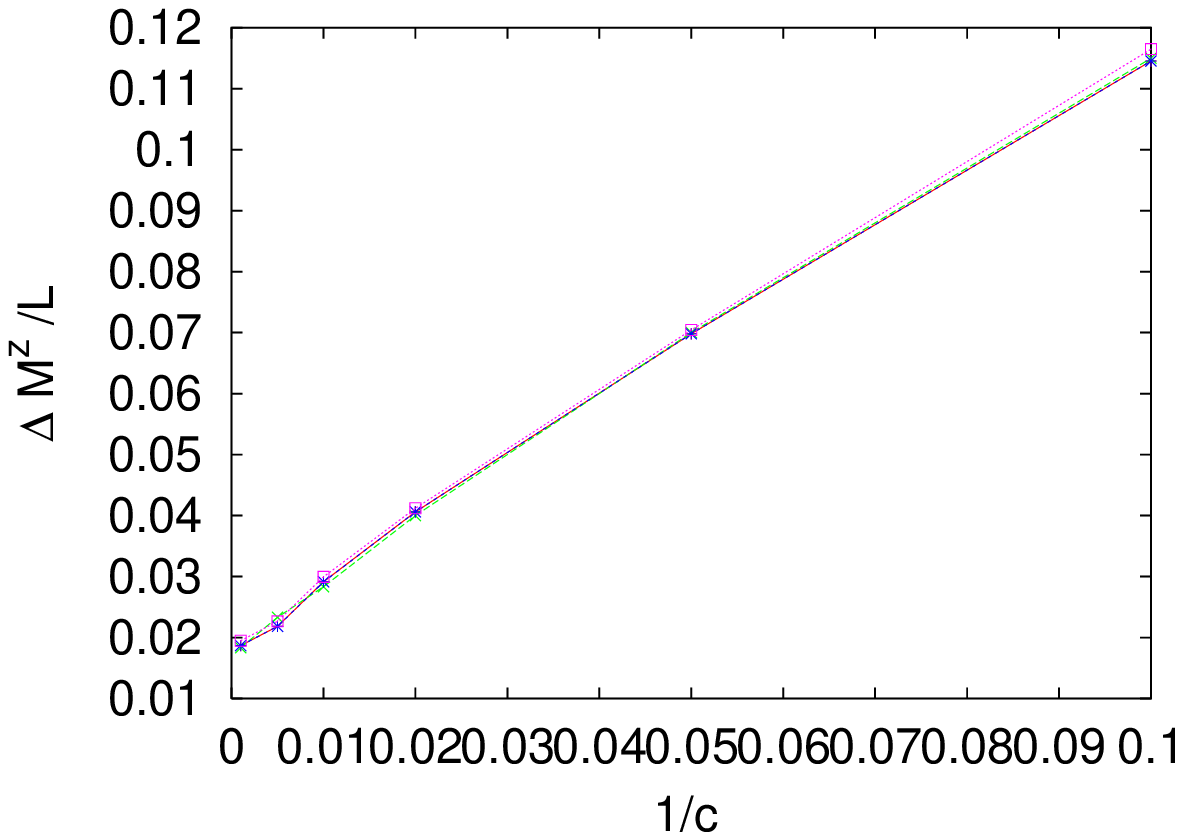}
}
\caption{
Left: The magnetization $M^z(t)$ as a function $H_z$ for $L=12$, $H_x=0.7$ 
and various sweep velocities.
Solid (red) line: $c=10$;
Long dashed (green) line: $c=20$;
Dashed (magenta) line: $c=50$;
Dotted (dark blue) line: $c=100$;
Dashed dotted (magenta) line: $c=200$.
Right: $\Delta M/ L$ as a function of $1/c$ for $Hx=0.7$ and various system sizes.
Solid (red) line: $L=4$;
Long dashed (green) line: $L=6$;
Dashed (dark blue) line: $L=10$;
Dotted (magenta) line: $L=12$. 
}
\label{fastsweep}
\end{center}
\end{figure}

Let us now study the behavior if we sweep much faster.
In Fig.~\ref{fastsweep}(left), we show the magnetization as a function of 
$t$ (or $H_z(t)$) for $L=12$ with $c=10,20,50,100$ and 200.
For these parameters, the data for other $L$ are almost indistinguishable 
from the $L=12$ data and are therefore not shown.
As Fig.~\ref{fastsweep}(left) shows, the magnetization 
oscillates about a stationary value for large values of $H_z$ where
the energy levels with different magnetization $M$ are far separated in the
energy-level diagram as we saw in Fig.~\ref{engHx07L12}.
Let us study the $c$-dependence of the saturated value $M_S=M_S(c)$.
In Fig.~\ref{fastsweep}(right), we plot the change of the magnetization
$\Delta M/L = (M_S(c)-(-L))/L$ as a function of $1/c$.
As shown in Fig.~\ref{fastsweep}(right), the data can be fitted well by
the expression
\beq
{\Delta M\over L}\simeq {\Delta M_0\over L} +{a\over c},
\eeq
where $\Delta M_0/L$ and $a$ are constants. 
These constants, to good approximation, do not depend on the system size.

In order to explain the observed $1/c$ dependence,
we introduce a perturbation scheme for fast sweeps (see Appendix B).
We regard the sweeping field (Zeeman) term as the zero-th order system and
treat the interaction among spins as the perturbation term.
The result is a series expansion in terms of $H_0/c$ (see Eq.~(\ref{Wperturb})), 
which explains the $1/c$ dependence.

In Appendix B, we also introduce a perturbation scheme based on independent
Landau-Zener systems each of which is given by a spin in a
transverse field 
with a sweeping field. 
In Appendix B, we show that this scheme can explain
the behavior of the magnetization dynamics in the fast-sweep regime.

\section{Summary and Discussion}

We have studied time-evolution of the magnetization in the ordered phase 
of the transverse Ising model under sweeping field $H_z$.
We found significant jumps of the magnetization 
at a certain value of the magnetic field which we called quantum spinodal point 
$H_z^{\rm SP}$.  
Although the energy-level diagram of the system significantly changes with the system size, 
we found size-independent magnetization processes for each pair $(H_x, c)$. 

In principle, it should be possible to understand the quantum dynamics of the magnetization
from the energy-level diagram of the total system. Indeed the picture of 
successive Landau-Zener scattering processes works in slow sweeping case~\cite{LZ1997}.
However, for fast sweeps, the time evolution can be regarded as an assembly
of local processes, the interaction between the spins being a perturbation.
Hence the dynamics of the magnetization does not depend on the size. 

When the quantum fluctuations are weak (small $H_x$), a series of local spin flips
governs the magnetization dynamics. 
The jumps of magnetization can be understood on the basis of
energy-level crossings of certain spin clusters (Eq.~(\ref{flipm})). 
The energy-level structure corresponding to the local cluster flips is, of course,
present in the energy-level diagram of the total system but
it is hidden in the complicated structure of the huge number of energy levels.

For large values of $H_x$ and fast sweeps, the magnetization process is also size-independent.
To explain this feature, we have introduced a perturbation scheme 
in which the small parameter is $H_0/c$.
In addition, we introduced a new perturbation scheme based of 
single-spin free Landau-Zener processes, which all together
have allowed us to provide an understanding of the magnetization dynamics 
under field sweeps in terms of the energy-level structure.

\section*{Acknowledgments}
\label{sec:ACK}

This work was partially supported by a Grant-in-Aid for Scientific 
Research on Priority Areas ``Physics of new quantum phases in superclean materials" 
(Grant No.\ 17071011), 
and also by the Next Generation Super Computer Project, 
Nanoscience Program of MEXT.
Numerical calculations were done on the supercomputer of ISSP. 

\appendix
\section{Size dependence of the energy gap at $H_z=0$}

\begin{figure}
\begin{center}
\mbox{
\includegraphics[clip,width=8cm]{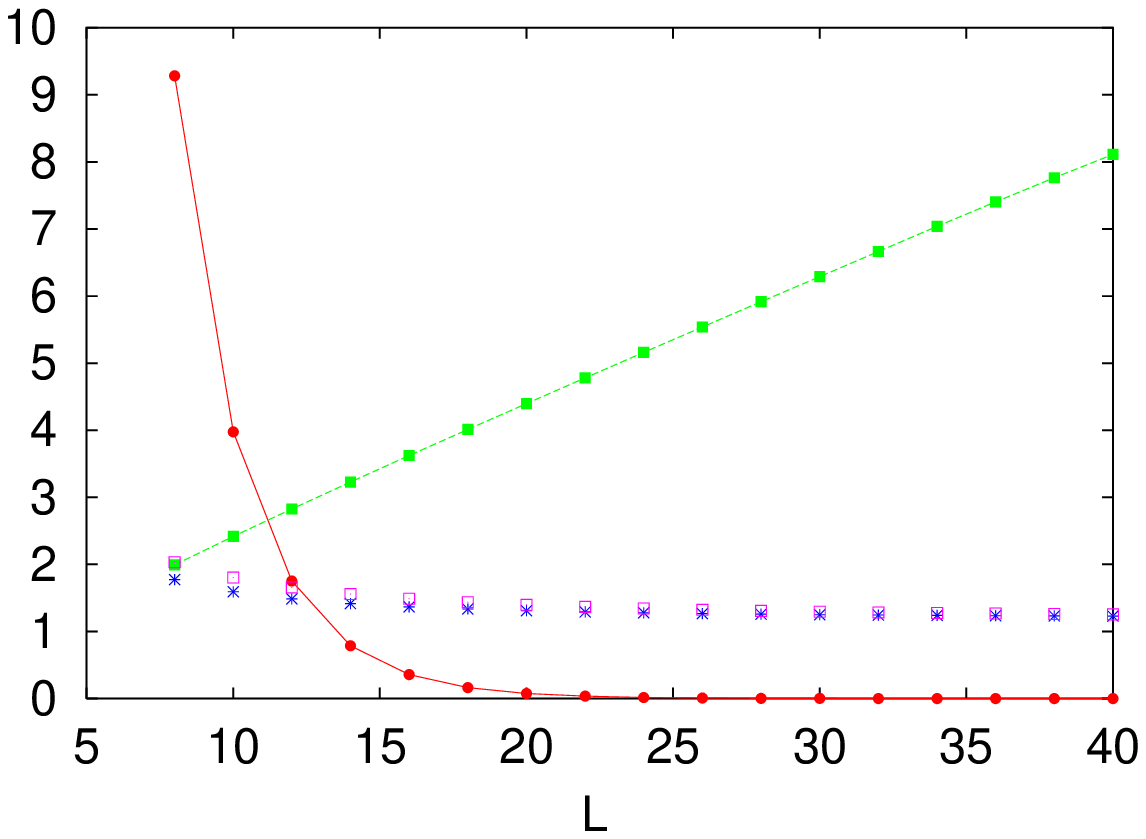}
\includegraphics[clip,width=8cm]{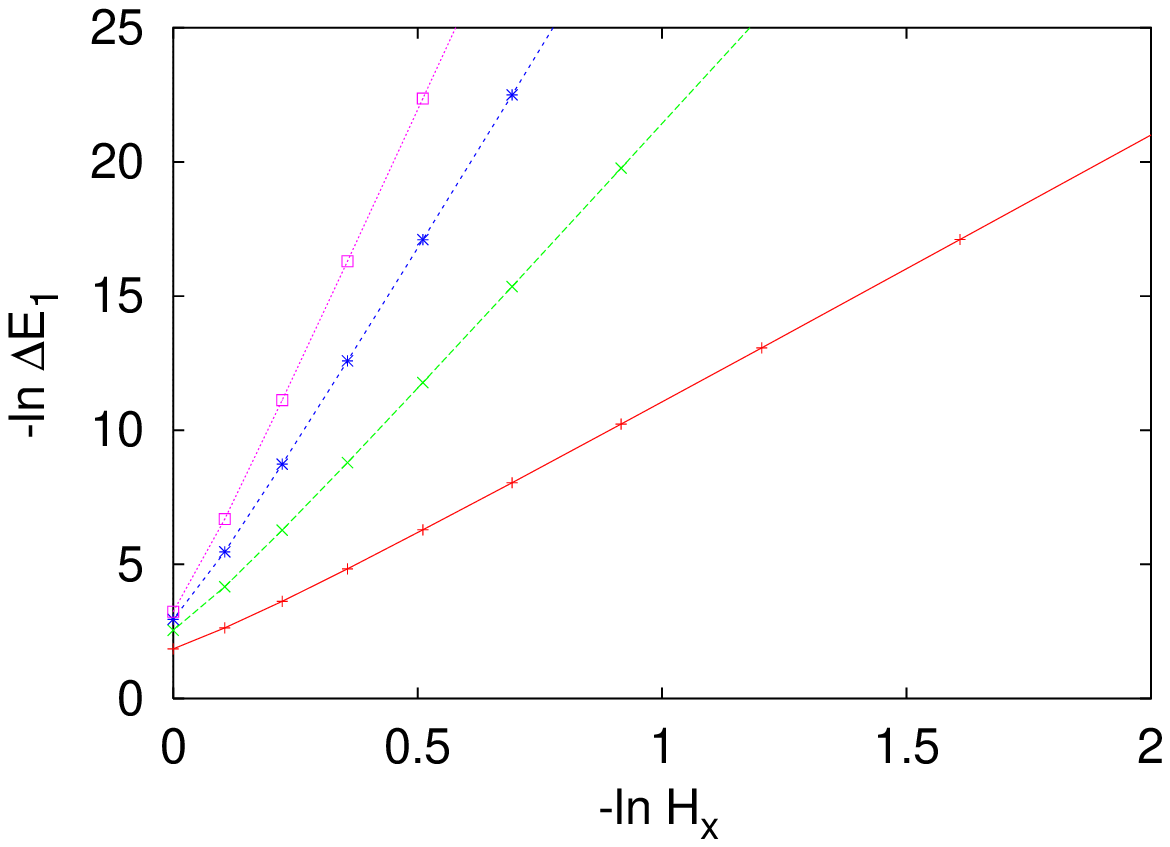}
}
\caption{(color online)
Left:
Size-dependence of the energy gaps for $H_x=0.7$. 
Bullets: Difference $500\times\Delta E_1/J$ between the energy of the first excited state
and the ground state energy.
This difference vanishes exponentially with $L$.
Solid squares: $-(\ln\Delta E_1/J)/2$;
Stars: $\Delta E_2/J$;
Open squares: $\Delta E_3/J$.
Right: The energy gap $\Delta E_1/J$ as a function
of $H_x$ for several $L$.
Plusses: $L=10$.
Crosses: $L=20$;
Stars: $L=30$;
Open squares: $L=40$.
Note the double logarithmic scale.
In both figures, lines are guides to the eyes only.
}
\label{energy-gap}
\end{center}
\end{figure}

The eigenvalues of the model Eq.~(\ref{ham}) are given by\cite{TI}
\beq
E=H_x\sum_q\omega_q\left(2\eta^{\dagger}\eta-1\right),
\eeq
where $\eta$ and $\eta^{\dagger}$ are fermion annihilation and creation operators, 
respectively, and
\beq
\omega_q=2\sqrt{1+2\lambda\cos q+\lambda^2},
\eeq
where $\lambda=J/H_x$.
When the number of the fermions is even, $q$ takes the values
\beq
q=\pm{\pi\over L}, \pm{3\pi\over L},\cdots, \pm{\pi(L-1)\over L},
\eeq
and when the number of the fermions is odd
\beq
q=0,\pm{2\pi\over L}, \pm{4\pi\over L},\cdots,\pm{\pi(L-2)\over L}, \pi.
\eeq
Because $\omega_q >0$, the ground state is given by $\eta^{\dagger}\eta =0$.
Thus, in the case of even number of fermions, the ground state is given by
\beq
E_{\rm E1} = -2H_x\sum_{m=1}^{L/2}
\sqrt{1+2\lambda\cos\left({L-2m+1\over L}\right)+\lambda^2},
\eeq
and the first excited state is given by
\beq
E_{\rm E2} = E_{\rm E1}
+4H_x\sqrt{1+2\lambda\cos\left({L-1\over L}\right)+\lambda^2}.
\eeq
In the case of an odd number of fermions, the lowest energy state is
\begin{eqnarray}
E_{\rm O1} &=& H_x\sqrt{1+2\lambda+\lambda^2}+H_x\sqrt{1-2\lambda+\lambda^2}
\nonumber \\
&&-2H_x\sum_{m=1}^{L/2-1}\sqrt{1+2\lambda\cos\left({L-2m+1\over L}\right)+\lambda^2},
\label{bel5h1n1}
\end{eqnarray}
and 
the first excited state is given by
\beq
E_{\rm O2} = E_{\rm O1}
+2H_x\sqrt{1-2\lambda+\lambda^2}+\sqrt{1+2\lambda\cos\left({L-2\over L}\right)+\lambda^2}.
\eeq
The energy gaps are given by
\beq
\Delta E_1= E_{\rm O1}- E_{\rm E1},
\eeq
\beq
\Delta E_2= E_{\rm E2}- E_{\rm E1},
\eeq
and 
\beq
\Delta E_3= E_{\rm O2}- E_{\rm E1}.
\eeq
Using these formulae, we can calculate the $L$-dependence of the gaps.
The results are plotted in Fig.~\ref{energy-gap}(left).
We find that $\Delta E_1$ vanishes exponentially with $L$,
that is
\beq
\Delta E_1 \propto \exp(-aL), 
\label{expdep}
\eeq
where $a$ depends on $\lambda$.
We also plot $-\log \Delta E_1$/2 to confirm
the exponential dependence. 
On the other hand, we find that $\Delta E_2$ is almost independent of $L$, and
$\Delta E_3$ is very close to $\Delta E_2$,
reflecting the fact that above the 3rd level 
the infinite system has a continuous spectrum.

It is also of interest to study the dependence of the energy gap 
$\Delta E_1$ on $H_x$ for several $L$. 
In Fig.~\ref{energy-gap}(right) we show the data
on double logarithmic scale.
In the regime of small $H_x$ we find a linear dependence
on $H_x$, suggesting that
\beq
\Delta E_1 \propto H_x^{2S}.
\eeq
Indeed, for small $H_x$ the slopes of the lines is given by $2S=L$.
This dependence on $H_x$ and $L$ is to be expected
when $L$ spins flip simultaneously.

\section{Perturbation analysis for Landau-Zener type sweeping processes}

When the sweep velocity $c$ is very large, the duration of the sweep is very short.
This suggests that it may be useful to
study the magnetization processes by a perturbational method in terms of 
the small parameter $1/c$.

Let us consider the following model.
\beq
{\cal H}={\cal H}_0+ctV,
\eeq
where ${\cal H}_0$ and $V$ are time independent.
We will work in the interaction representation with respect to $ctV$,
that is, we take the motion of $ctV$ as reference,
not ${\cal H}_0$ as is usual done.
The Schr\"odinger equation is
\beq
i\hbar\pd |\Psi\rangle=\left({\cal H}_0+ctV\right) |\Psi\rangle.
\eeq
In the interaction representation we have
\beq
|\Psi\rangle=e^{-ict^2V/2\hbar}|\Phi\rangle,
\eeq
and the equation of motion is given by
\beq
i\hbar\pd |\Psi\rangle=
i\hbar (-ictV/\hbar)e^{-ict^2V/2\hbar}|\Phi\rangle
+i\hbar e^{-ict^2V/2\hbar}\pd |\Phi\rangle
=e^{-ict^2V/2\hbar}\left(ctV+i\hbar \pd \right)|\Phi\rangle,
\eeq
and therefore the Schr\"odinger equation for $|\Phi\rangle$ is given by
\beq
i\hbar\pd |\Phi\rangle=e^{ict^2V/2\hbar}{\cal H}_0e^{-ict^2V/2\hbar}|\Phi\rangle.
\eeq
Defining
\beq
W(t)\equiv e^{ict^2V/2\hbar}{\cal H}_0e^{-ict^2V/2\hbar},
\eeq
we can use the usual perturbation expansion scheme for
\beq
i\hbar\pd |\Phi\rangle=W(t) |\Phi\rangle,
\eeq
and find
\beq
|\Phi(t)\rangle=\left[1+\left(1\over i\hbar\right)\int_{t_0}^tW(t_1)dt
+\left(1\over i\hbar\right)^2
\int_{t_0}^t\int_{t_0}^{t_1}W(t_1)W(t_2)dt_1dt_2+\cdots\right]|\Phi(0)\rangle.
\label{Wperturb}
\eeq
In the sweep ($-H_0< ct < H_0$), $t_0=-{H_0/c}$ and $t={H_0/c}$.
Thus, the integral is of order $H_0/c$. Therefore we can regard the above 
expansion is a series expansion in terms of power of $H_0/c$.
Of course the series can be also regarded as a power of ${\cal H}_0$ as
in the usual sense.

\subsection{Transverse Ising model under a field sweep}
Now, we consider our problem
\beq
{\cal H}(t)=-J\sum_j\sigma_j^z\sigma_{j+1}^z-H_x\sum_j\sigma_j^x
-ct\sum_j\sigma_j^z.
\eeq
We set
\beq
{\cal H}_0=-J\sum_j\sigma_j^z\sigma_{j+1}^z-H_x\sum_j\sigma_j^x
\eeq
and
\beq
V=-\sum_j\sigma_j^z.
\eeq
Then, $W(t)$ is given by
\begin{eqnarray}
W(t)&=&e^{-ict^2/2\hbar\sum_j\sigma_j^z}
{\cal H}_0e^{ict^2/2\hbar\sum_j\sigma_j^z}
\nonumber \\
&=&-J\sum_j\sigma_j^z\sigma_{j+1}^z
-H_x\sum_j\left(\sigma_j^+e^{-ict^2/\hbar}+\sigma_j^-e^{ict^2/\hbar}\right).
\end{eqnarray}
We may include the diagonal term $-J\sum_j\sigma_j^z\sigma_{j+1}^z$ in $V$.
Then the expansion is regarded as series of $H_x$.
This expansion corresponds to the series of jumps discussed in Eq.~(\ref{flipm}).

We also note that if $J=0$ the above process is an ensemble of 
independent Landau-Zener processes. 
Each of them is independently expressed by
\beq
i\hbar\pd|\Phi_{\rm LZ} (t)\rangle=-H_x\left(\sigma^+e^{-ict^2/\hbar}
+\sigma^-e^{ict^2/\hbar}\right)|\Phi_{\rm LZ} (t)\rangle.
\eeq

\subsection{Perturbation theory in terms of independent LZ systems}

Next, we consider the case in which the transverse field 
is included in $V$. We sweep the field from $-H_0$ to $H_0$. 
The duration of the sweep is ${2H_0/c}$.
We assume that
\beq
H_0\gg J > H_x,
\eeq
such that the motion due to $V$ is that of an ensemble of 
independent Landau-Zener processes.
Thus, we consider the ensemble of the LZ systems as the unperturbed system.

We know the properties of each system. Namely, we know that the scattering 
becomes small when $c$ becomes large.
The time evolution of each LZ system is given by
\beq
\left(\begin{array}{c} 1 \\ 0 \end{array}\right) 
\rightarrow
e^{i\phi(t)} \left(\begin{array}{c} \sqrt{p} \\
\sqrt{1-p}e^{i\varepsilon(t)}\end{array}\right)
\equiv\psi(t),
\label{LZscatter}
\eeq
in the adiabatic basis, that is in the representation 
that uses the eigenstates of the system with given $H_z(t)$. 
Here, $p$ is the probability for staying the ground state. 
In the Landau-Zener theory, $p$ is given by the well-known
expression 
\beq
p=1-\exp\left(-{\pi H_x^2\over\hbar c}\right).
\label{LZp}
\eeq
In the case of small $H_0$, $p$ may have a different form.
Even in those cases, the expression Eq.~(\ref{LZscatter}) is still correct
and the present formulation works if we employ a correct expression for $p$.

\begin{figure}
\begin{center}
\mbox{
\includegraphics[clip,width=8cm]{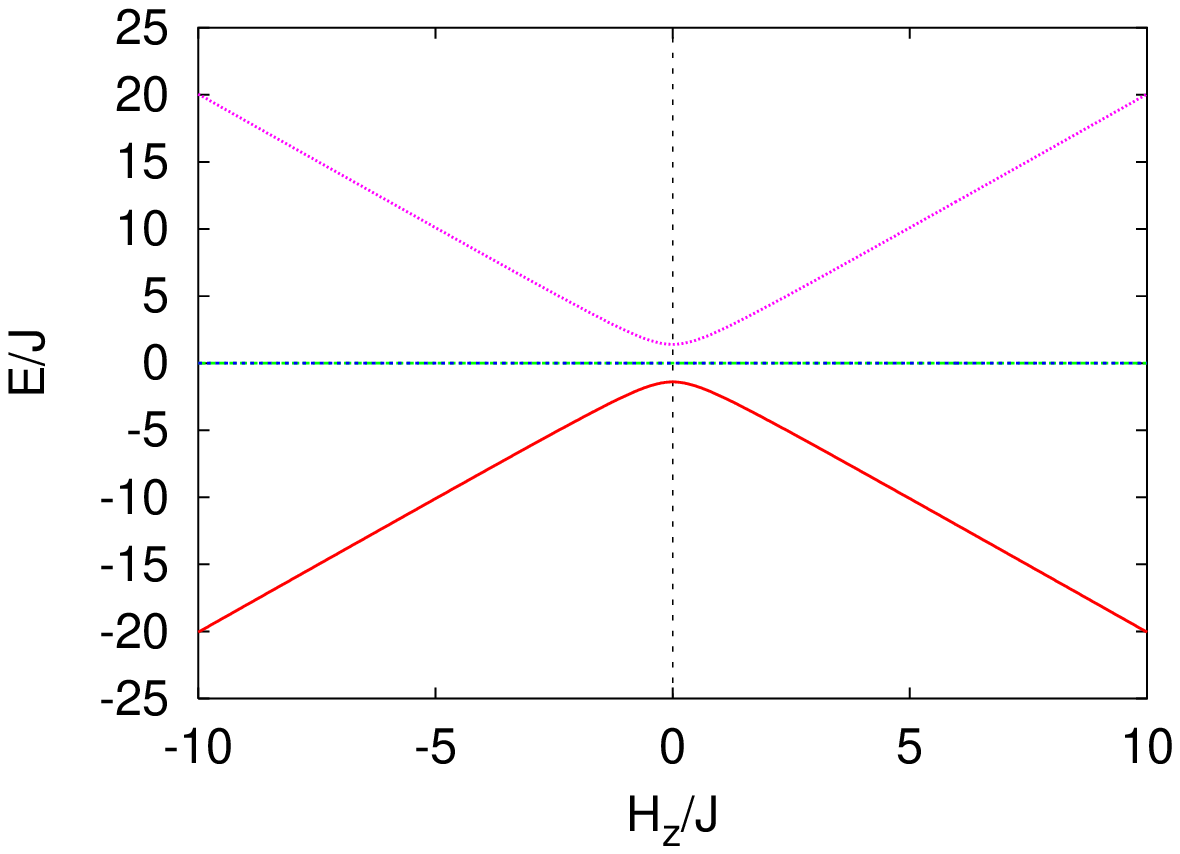}
\includegraphics[clip,width=8cm]{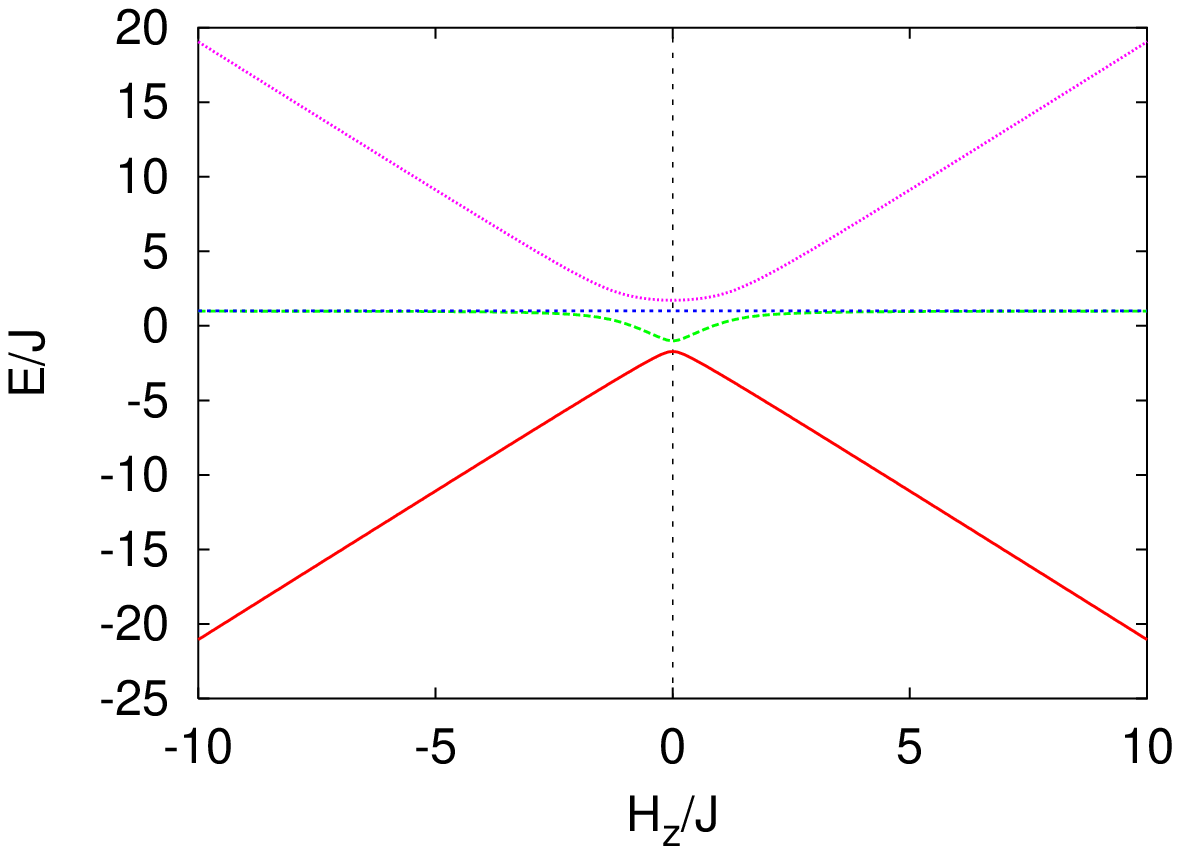}
}
\caption{Energy level diagram for a two-spin Landau-Zener model Eq.~(\ref{LZ2}) with $H_x=0.7$.
Left: $J=0$;
Right: $J=1$.
}
\label{Energy1}
\end{center}
\end{figure}

The unperturbed state is given by
\beq
\Phi_0(t)=\prod_j\psi_j(t)=e^{iL\phi(t)}
\left(\begin{array}{c} \sqrt{p} \\\sqrt{1-p}e^{i\varepsilon(t)}\end{array}\right)_1\otimes
\left(\begin{array}{c} \sqrt{p} \\\sqrt{1-p}e^{i\varepsilon(t)}\end{array}\right)_2\otimes\cdots\otimes
\left(\begin{array}{c} \sqrt{p} \\\sqrt{1-p}e^{i\varepsilon(t)}\end{array}\right)_L.
\eeq
The zero-th order is given a usual Landau-Zener process of which the energy diagram 
is given by Fig.~\ref{Energy1}(left), which shows
the energy-level diagram for the two independent spins. 
Thus, in this case there are four states, $ (++),(+-), (-+),$ and $(--)$. 
The states consisting of  $ (+-)$ and $ (-+)$ are degenerate with energy zero.

The interaction term $-J\sum_j\sigma_j^z\sigma_{j+1}^z$ is the perturbation.
As far as the expansion Eq.~(\ref{Wperturb}) converges
with less than $L$-th terms, it gives a local effect.
To first order in $J$, only the nearest-neighbor spins interact, 
giving a contribution of the order $J$.
The sweep-velocity dependence is taken into account through the zero-th order term.
If we take a large $H_0$, the integral in Eq.~(\ref{Wperturb}) is no longer
small, and we have to regard Eq.~(\ref{Wperturb}) as a series of $J$.
Therefore, we do not have any small parameter, and the Eq.~(\ref{Wperturb})
represents the original general dynamics.
In the case of fast sweeps, the effective range of quantum mixing in which
the diabatic energy levels (levels for $H_x=0$) cross each other, is of order
$L\times J$, and therefore the duration of interaction is of order $LJ/c$.
Hence, the integration gives a contribution of order $LJ/c$ which now becomes
the small parameter.
In the case of finite $H_0$, the small parameter is the minimum of
$(H_0/c, LJ/c)$. In the present study, $H_0=1$. Then $H_0/c$ is the small
parameter and we cannot use the form of $p$ given in Eq.~(\ref{LZp}).
In any case, the series converges for the fast sweeps and we expect that
the perturbation effect does not depend on $L$.

\begin{figure}
\begin{center}
\mbox{
\includegraphics[clip,width=8cm]{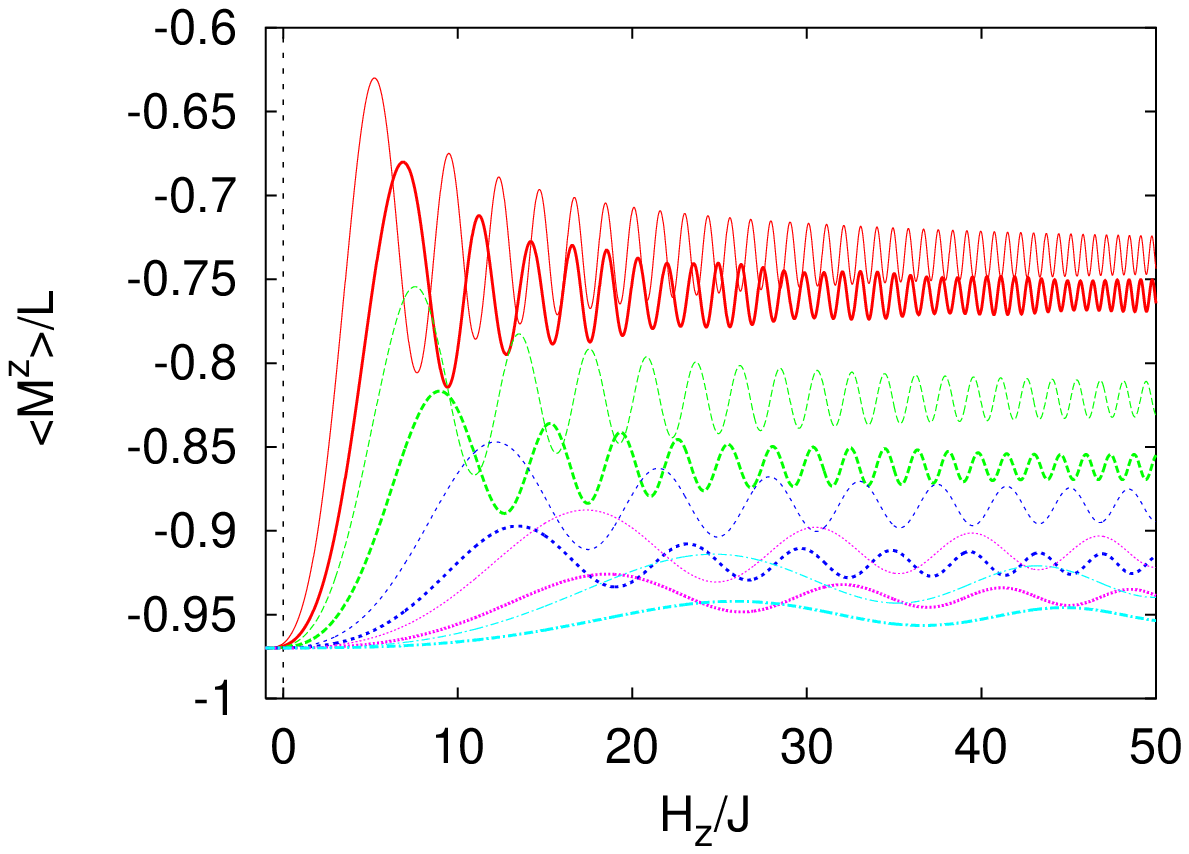}
\includegraphics[clip,width=8cm]{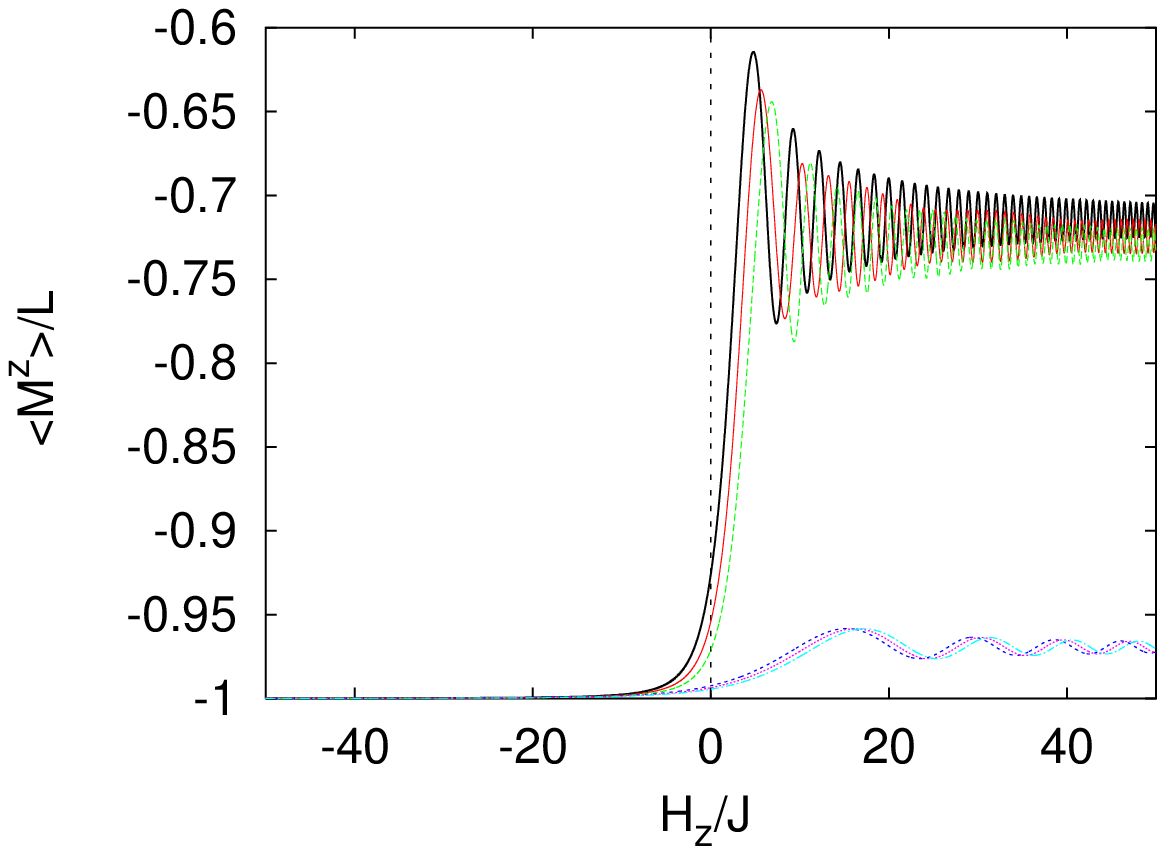}
} 
\caption{
Left: Comparison of the magnetization processes of the model Eq.~(\ref{LZ2}) 
with $J=0$ (thin line)
and $J=1$ (thick lines)
for $H_x=0.7$. 
Solid (red) line: $c=10$;
Long dashed (green) line: $c=20$;
Dashed (magenta) line: $c=50$;
Dotted (dark blue) line: $c=100$;
Dashed dotted (magenta) line: $c=200$.
For each $c$, the magnetization of the single LZ process
is shifted by an amount such that at $H_z=-1$
it coincides with the magnetization of the model Eq.~(\ref{LZ2}) with $J$ replaced by $2J$.
Right: Comparison of the magnetization of a single LZ process,
that of the model Eq.~(\ref{LZ2}), and that of the model Eq.~(\ref{LZ2}) with $J$ replaced by $2J$. 
$H_x=0.7$ and $c=100$. 
Solid (black) line: $c=10$, single LZ process;
Solid (red) line: $c=10$, Eq.~(\ref{LZ2});
Long dashed (green) line: $c=10$, Eq.~(\ref{LZ2}) with $J$ replaced by $2J$;
Dashed (magenta) line: $c=100$, single LZ process;
Dotted (red) line: $c=100$, Eq.~(\ref{LZ2});
Dashed dotted (blue) line: $c=100$, Eq.~(\ref{LZ2}) with $J$ replaced by $2J$.
}
\label{LZ1-LZ2J1}
\end{center}
\end{figure}

The system described by the first-order perturbation theory corresponds to a Hamiltonian 
of two spins exhibiting the Landau-Zener scattering process 
and which are coupled by an Ising interaction. The Hamiltonian reads
\beq
{\cal H}_{\rm CLZ}=-J\sigma_1^z\sigma_2^z
-\left(H_x\sigma_1^x-ct\sigma_1^z\right)
-\left(H_x\sigma_2^x-ct\sigma_2^z\right).
\label{LZ2}
\eeq
The energy-level diagram of this system is shown in Fig.~\ref{Energy1}(right).
Let us study 
the effect of the interaction on the dynamics in this case.
We compare the magnetization processes of the model Eq.~(\ref{LZ2})
with $J=0$ and $J=1$.
The results are shown in Fig.~\ref{LZ1-LZ2J1}(left).
Note that the sweep starts from $H_z=-H_0=-1$.

Next, in Fig.~\ref{LZ1-LZ2J1}(right),
we show the magnetization processes for $c=100$
for the model Eq.~(\ref{LZ2}) with that 
of the same model with $J$ replaced by $2J$.
If we use a small value of $H_0$, the ground states
of the models at $H_z=-1$ differ significantly. 
Therefore, to compare the results, in this figure,
we take $H_0=-60$ such that the ground state of both models is
close to the all-spins-down state. 
The average of the first and the third curves is close to the second curve. 
This fact indicates that the processes are well described by
the first-order perturbation theory. Indeed, the deviation from the single Landau-Zener
model is 0, $J$, and $2J$, respectively.

\begin{figure}
\begin{center}
\mbox{
\includegraphics[clip,width=8cm]{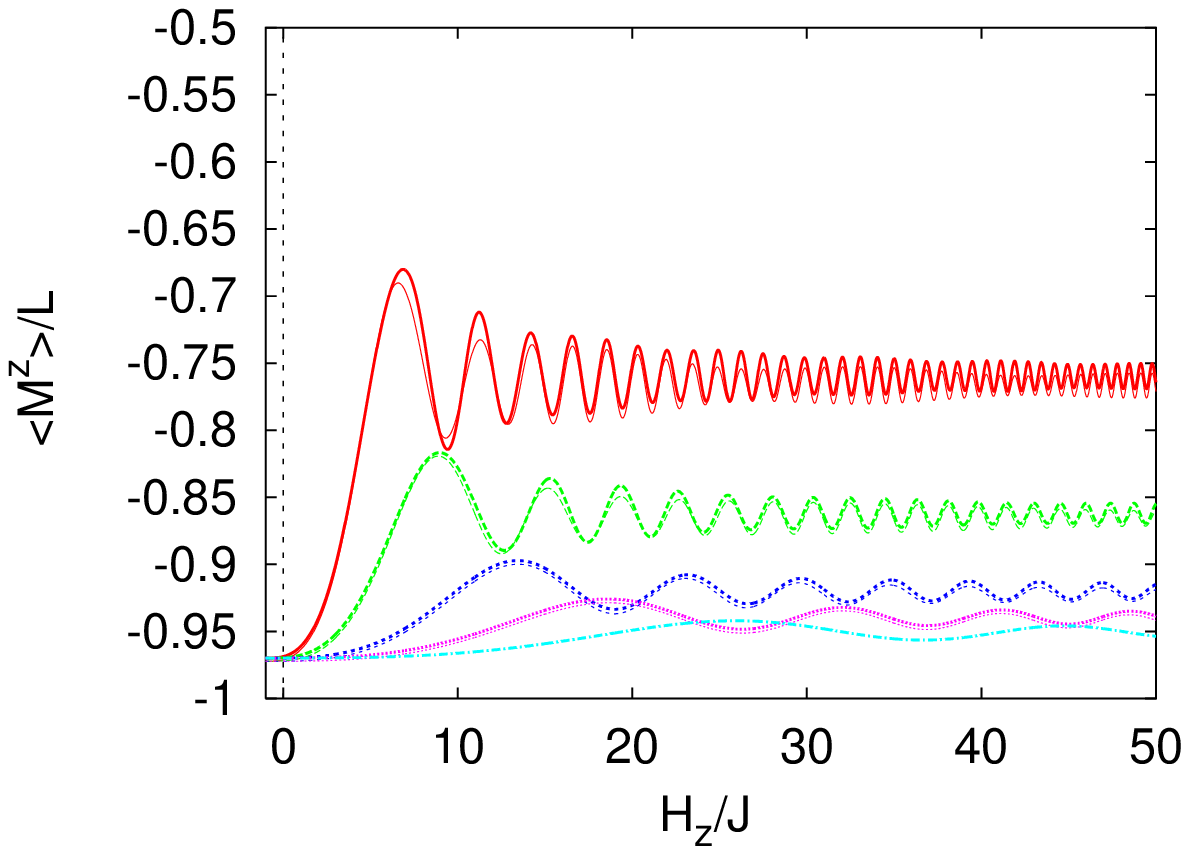}
\includegraphics[clip,width=8cm]{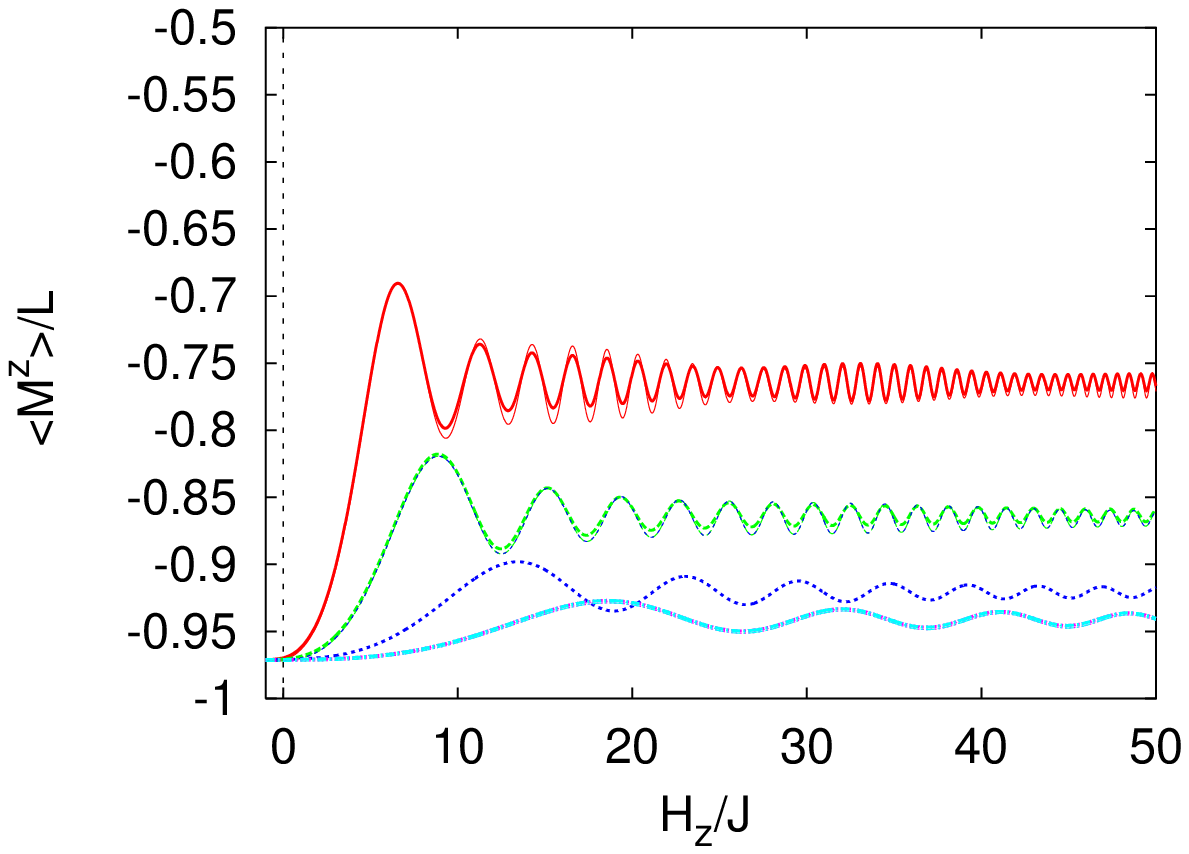}
}
\caption{
Left:
Comparison of the magnetization processes of the model Eq.~(\ref{LZ2}) with $J$ replaced by $2J$ (thin lines)
and the model with three spins (thick lines) for $H_x=0.7$.
Solid (red) line: $c=10$;
Long dashed (green) line: $c=20$;
Dashed (magenta) line: $c=50$;
Dotted (dark blue) line: $c=100$;
Dashed dotted (magenta) line: $c=200$.
Right: Same as left except that the comparison is between
models with $L=4$ (thin lines) and $L=12$ (thick lines) for $H_x=0.7$.
}
\label{LZ2-LZ3L12}
\end{center}
\end{figure}

We also compare the magnetization processes of the models Eq.~(\ref{LZ2}) 
with $J$ replaced by $2J$ and that of a model with 3 spins 
in Fig.~\ref{LZ2-LZ3L12}(a).
The difference between the models of 4 spins and of 12 spins 
is also shown in Fig.~\ref{LZ2-LZ3L12}(right).
In all these cases, we start at $H_z=-1$ because the 
magnetizations per spin are very close in all the cases.
We find almost no difference, indicating that the processes are well
described by the first-order perturbation theory.

When the sweep velocity becomes small, we may need higher order perturbation terms.
If the relevant order of the perturbation is less than the length of the chain,
we expect a size-independent magnetization process.
The size independent magnetization in the quantum spinodal decomposition 
can be understood in this way.

The local motion of magnetization can be understood from a view point of  
an effective field from the neighboring spins.
We may study the magnetization process of a single-spin in
a dynamical mean-field generated by its neighbors.\cite{Hams}
Let us describe the situation by the following Hamitonian:
\beq
{\cal H}_{\rm MF}= -(H_z(t)+2J\langle \sigma^z\rangle)\sigma^z-H_x\sigma^x.
\label{Hmean}
\eeq
Because the mean field is almost $2J$ during the fast sweep,
the mean field simply shifts $H_0$ by a constant $2J$.
Thus, we conclude that for fast sweeps,
the dynamics is very similar to that of a single spin, meaning that
for the dynamics, the effective field on each spin in the lattice is essentially
that same as the applied field. This conclusion is consistent with our earlier
comparison of the zero-th and first-order perturbation results.

\end{document}